\documentclass[superscriptaddress,onecolumn, reprint ]{revtex4-2}
\usepackage{graphicx}
\usepackage{amssymb}
\usepackage{epstopdf}
\usepackage{amsmath,dsfont, commath}
\usepackage{bm}
\usepackage{braket}
\usepackage{changes}
\usepackage{hyperref}
\newcommand{\be}{\begin{equation}}
	\newcommand{\ee}{\end{equation}}
\newcommand{\bea}{\begin{eqnarray}}
	\newcommand{\eea}{\end{eqnarray}}

\newcommand{\ba}{\begin{array}}
	\newcommand{\ea}{\end{array}}

\newcommand{\bl}{\begin{flalign}}
	\newcommand{\enl}{\end{flalign}}

\newcommand{\pa}{\partial}
\newcommand{\mc}[1]{\mathcal{#1}}

\newcommand{\tdse}{time dependent Schr\"{o}dinger equation}

\newcommand{\eq}[1]{Eq. \eqref{#1}}

\newcommand{\fig}[1]{Fig. (\ref{#1})}

\usepackage{cleveref}% load last

\newcommand{\half}{\frac{1}{2}}

\newcommand{\proj}[1]{\ket{#1}\bra{#1}}

\renewcommand{\bf}[1]{\mathbf{#1}}
\newcommand{\grad}{\nabla}

\begin{document}

	%opening
	\title{Local diabatic representation of conical intersection quantum dynamics }%coupled electron-nuclear motion}
	\author{Bing Gu}

	\email{gubing@westlake.edu.cn}
	\affiliation{Department of Chemistry, School of Science, Westlake University, Hangzhou, Zhejiang 310030, China}
\affiliation{Institute of Natural Sciences, Westlake Institute for Advanced Study, Hangzhou, Zhejiang 310024, China}
		\begin{abstract}
		%Electronic coherence 
		Conical intersections are ubiquitous in polyatomic molecules and responsible for a wide range of phenomena in chemistry and physics. 
		We introduce and implement a local diabatic representation for the correlated electron-nuclear dynamics around conical intersections.  It employs the adiabatic electronic states but avoids the singularity of nonadiabatic couplings, and is robust to different gauge choices of the  electronic wavefunction phases. Illustrated by a two-dimensional conical intersection model, this representation captures nonadiabatic transitions, electronic coherence, and geometric phase. %It does not require force evaluations. 
	%	Thus, this representation paves the way for ab initio conical intersection dynamics directly with adiabatic electronic states.  
	\end{abstract}
	\maketitle

\section{Introduction}

Conical intersection (CI), degeneracy points in the adiabatic potential energy surfaces, is   ubiquitous in polyatomic molecules.  It is responsible for  virtually all photochemical  and photophysical processes including nonradiative relaxation, Jahn-Teller effect, vision, photo-stability of DNA molecules \cite{longuet-higgins1958, domcke2011, yarkony1996, baer2006, improta2016}.  In the vicinity of CIs, the electronic and nuclear motion becomes strong coupled,  thus the adiabatic Born-Oppenheimer approximation breaks down. An immediate consequence of the strong electron-nuclear (vibronic) coupling is the nonradiative electronic relaxation.  Furthermore, transient electronic coherences may emerge during the passage through a CI. It has been suggested that such transient coherence can be probed by stimulated x-ray Raman scattering \cite{kowalewski2015b, keefer2020} and twisted x-ray diffraction \cite{yong2022a}, which can provide a direct spectroscopic signature of CIs. 

Another  effect due to CI is geometric phase \cite{wittig2012, berry1998, longuet-higgins1958}. A nuclear trajectory encircling a CI in the configuration space will pick up a geometric phase \cite{berry1984}.  The geometric phase arises not only in the nonadiabatic dynamics, but also  in the adiabatic dynamics even when the CI is energetically inaccessible. This has been observed experimentally in the H + HD $\rightarrow$ H2 + D reaction \cite{yuan2018a}. Incorporating the geometric phase into the molecular dynamics is essential to understand the adiabatic and nonadiabatic molecular dynamics \cite{ryabinkin2017}. 

Understanding the CI dynamics requires solving the nuclear wave packet dynamics in multiple electronic potential energy surfaces \cite{curchod2018}. This is usually done with exact quantum dynamics approaches in ab initio computed potential energy surfaces \cite{guo2016}. 
 Approximate methods with a classical treatment of nuclei including the widely used trajectory surface-hopping \cite{Tully1990} and Ehrenfest dynamics \cite{li2005a}, being useful to describe electronic relaxation dynamics in large-scale molecular systems, cannot describe properly the transient electronic coherence and geometric phase \cite{guo2016, ryabinkin2014}. 
The nonadiabatic conical intersection dynamics is either performed in the adiabatic or the diabatic representation. In the adiabatic representation, the Born-Huang ansatz is used for the molecular wavefunction, $\Psi(\bf r, \bf R, t) = \sum_\alpha {\phi_\alpha(\bf r; \bf R)} \chi_\alpha(\bf R, t)$ where $\phi_\alpha(\bf r;\bf R)$ is the $\alpha$th adiabatic electronic state depending parametrically on the nuclear configuration $\bf R$ and $\chi_\alpha(\bf R, t)$ is the associated nuclear wave packet. This leads to the intuitive picture of each nuclear wave packet propagating in its own adiabatic potential energy surfaces (APES) and makes electronic transitions when it encounters a region in the configuration space where the nonadiabatic couplings becomes significant, e.g. close to a CI.  The gauge freedom in the Born-Huang approach is a local U(1) phase   transformation  ${\chi_\alpha'}(\bf R) = e^{i\theta_\alpha(\bf R)} {\chi_\alpha(\bf R)}, {\phi_\alpha'}(\bf r; \bf R) = e^{-i\theta_\alpha(\bf R)} {\phi_\alpha}(\bf r; \bf R)$.  In this representation, the nonadiabatic coupling accounts for all nonadiabatic effects. Note that the nuclear wave packets  are gauge-dependent and thus cannot be experimentally observed.  A major problem running quantum dynamics on the adiabatic potential energy surfaces is that in the presence of CI,  the nonadiabatic coupling becomes singular. This is because the nonadiabatic coupling depends inversely on the energy gap 
%$\bf d_{\alpha \beta} = \frac{1}{E_\alpha(\bf R) - E_\beta(\bf R)}$ 
and the gap vanishes at the CI point where two APESs intersect.

Transforming to the diabatic representation can avoid  such singularities, whereby the couplings between diabatic  states are well-behaved functions of nuclear coordinates. In fact, there are quantum dynamics methods that can only be used under the diabatic representation \cite{mandal2018a}. However, exact diabatization in general does not exist within a finite number of electronic states due to topological obstruction \cite{mead1982}. Various approximate quasi-diabatization methods  have been proposed \cite{yarkony2019, subotnik2008} based on different criteria. This may introduce spurious singularities or require further approximations \cite{yarkony2019}. For example, in the adiabatic-to-diabatic transformation approach, 
the residual couplings, the part of nonadiabatic couplings that cannot be transformed into diabatic couplings, are usually neglected \cite{Choi2021}.

%Here we propose a locally diabatic representation. 
%
%
%We first choose a basis set $\set{\chi_\mu(\bf R)}$ to describe the nuclear motion.   We construct the  localized basis functions by diagonalizing the position matrix elements $x_{\mu \nu} = \braket{\chi_\mu | x|\chi_\nu}$. This is  discrete variable representation \cite{light2000}. 
%% By construction, the basis set are the maximally localized states.  
%%$\ket{x_n}$ are eigenstates of position operator, and hence localized in configuration space. 
%The electronic Hamiltonian
%
%It follows that
%\be
%H_{BO}(\bf r, \bf R) \ket{\bf R_\alpha} = H_{BO}(\bf r; \bf R_n) \ket{\bf R_\alpha}
%\ee

Here we propose a locally diabatic representation (LDR) for the coupled electron-nuclear quantum dynamics at a CI. This representation avoids the singularity of nonadiabatic couplings because the nuclear kinetic energy operator does not operate on the electronic states. Moreover, while being a diabatic representation, it employs the adiabatic electronic eigenstates. Thus, it is straightforward to combine it with the well-established quantum chemistry methods, that usually computes the adiabatic electronic states at a fixed nuclear geometry. The LDR can be taken as a generalization of the crude adiabatic representation, wherein electronic states at a single reference nuclear configuration are chosen as the basis set\cite{tannor2007}. This representation is crude in the sense that the  chosen electronic basis will not be appropriate when the nuclear configuration deviates far from the reference geometry.  In the LDR, many reference nuclear configurations, determined by a discrete variable representation of the nuclear coordinates, are chosen. 
The nonadiabatic transitions and geometric phase effects are all included in an electronic overlap matrix, the overlap between electronic states at different nuclear geometries. This overlap matrix is always well-behaved,  even when the adiabatic electronic states are not smooth with respect to the nuclear coordinates. Therefore, the adiabatic states  obtained from electronic structure computations can be directly used without further smoothing procedure. 
An illustration of the utility of LDR is made for modeling the nonadiabatic wave packet dynamics of a two-dimensional CI model.

\section{Local Diabatic Representation}
The LDR is constructed as follows. In contrast to the adiabatic representation, we consider the nuclear motion first and then electronic. 
We  choose a primitive basis set $\set{\chi_\mu(\bf R)}$ of size $N$, not necessarily orthogonal, to describe the nuclear motion.  % The nuclear basis functions are not necessarily orthogonal, e.g., the Gaussian basis. 
This basis set should cover the relevant nuclear configuration space of the target process. 
We then construct the localized orthogonal basis functions by solving a generalized eigenvalue problem of the position operators. Since all position operators commute, they share common eigenstates labeled by  their eigenvalues $\ket{\bf R_n}$ with $n$ running over the eigenstates .
Expand
$\ket{\bf R_n} = \sum_\mu U_{\mu n} \ket{\chi_\mu}$, the transformation matrix can be obtained by solving the eigenvalue problem for each position operator $x$
\be 
X U = S U \Lambda,
\ee
$X_{\mu \nu} = \braket{\chi_\mu | x |\chi_\nu}$, $S_{\mu \nu} = \braket{\chi_\mu |\chi_\nu}$ is  the nuclear overlap matrix, and the diagonal matrix $\Lambda$ contains the position eigenvalues.  %and hence localized in configuration space. 
This basis set defines a resolution of identity of the nuclear space $I = \sum_{\mu, \nu} \ket{\chi_\mu} \del{S^{-1}}_{\mu \nu} \bra{\chi_\nu} = \sum_n \proj{\bf R_n}$.

%It follows that
%\be
%H_{BO}(\bf r, \bf R) \ket{\bf R_\alpha} = H_{BO}(\bf r; \bf R_n) \ket{\bf R_n} \label{eq:111}
%\ee where $H_\text{BO}$ is the electronic Hamiltonian. \eq{eq:111} is valid under the finite basis representation.

%The electronic structure computations then yield the adiabatic electronic eigensates, i.e., $H_\text{BO}(\bf R_n)\phi_\alpha(\bf r; \bf R_n) = E_\alpha(\bf R_n) \phi(\bf r; \bf R_n) $, where $\alpha$ labels the electronic states.

 Our ansatz for the full molecular wavefunction is now  given by
\be
\begin{split}
\Psi(\bf r, \bf R, t) &= \sum_n \Phi_n(\bf r, t; \bf R_n) \chi_n(\bf R)  \\
&= \sum_n {\sum_\alpha C_{n\alpha}(t) \phi_\alpha(\bf r; \bf R_n)} \chi_n(\bf R; \bf R_n)
\end{split}
\label{eq:100}
\ee
where $\chi_n(\bf R)= \braket{\bf R|\bf R_n}$ is a nuclear wave function centered at $\bf R_n$, $\phi_\alpha(\bf r, \bf R_n)$ is the $\alpha$th adiabatic electronic state $H_\text{BO}(\bf R_n)\phi_\alpha(\bf r; \bf R_n) = E_\alpha(\bf R_n) \phi(\bf r; \bf R_n) $ with $H_\text{BO}(\bf R) = H - \hat{T}_\text{n}$ the electronic Hamiltonian, the full molecular Hamiltonian subtracting the nuclear kinetic energy operator. 
Here $E_\alpha(\bf R)$ are the adiabatic potential energy surfaces.

In comparison to the adiabatic representation, the ansatz reverses the role of electrons and nuclei: instead of thinking about a time-dependent nuclear wave packet associated with a fixed APES, we consider a time-dependent electronic state $\Phi_n(\bf r, t; \bf R_n)$ associated with a fixed nuclear basis. In the second step of \eq{eq:100}, we have expanded the electronic state in terms of the adiabatic electronic basis at a reference geometry defined by the associated nuclear basis. 
This ansatz can then be understood as an expansion of the full molecular wavefunction in terms of the  electron-vibrational (vibronic) basis set  $\set{\phi_\alpha(\bf r; \bf R_n) \chi_n(\bf R) }$ with expansion coefficients $C_{n\alpha}(t)$. 
Following the orthonormality of the nuclear basis set, the vibronic basis set is also orthonormal in the joint electron-nuclear space, i.e., $\braket{\phi_\beta(\bf R_m), \bf R_m | \phi_\alpha(\bf R_n), \bf R_n } = \delta_{\beta \alpha}\delta_{nm}$. 

There is a gauge structure in \eq{eq:100}, 
\be 
\ket{\bf R_n'} = e^{i \theta_n} \ket{\bf R_n}, ~~~ \ket{\phi_\alpha'(\bf R_n)} = e^{-i\theta_n} \ket{\phi_\alpha(\bf R_n)}.
\label{eq:gauge}
\ee
This is a discretized form of the gauge transformation in the Born-Huang expansion, $U = e^{i \theta(\bf R)}$. However, while $\theta(\bf R)$ is required to be continuous function of $\bf R$ in the Born-Huang approach, the $\theta_n$ can be independently and arbitrarily chosen. This is  because, as will be shown below, our dynamical equation does not involve nuclear derivative of the electronic wavefunctions.

%if we use an ansatz
%\be
%\Psi(\bf r, \bf R, t) = \sum_n \del{\sum_\alpha C_{n\alpha}(t) \phi_\alpha(\bf r; \bf R_n)} \chi_n(\bf R; \bf R_n, \bf P_n)
%\ee
%where $\chi$ is a Gaussian wavepacket centered at $\bf R_n$.
%This avoids the singular NAC at CIs.

%\be H \ket{\phi} = E \ket{\phi}
%\ee
%
%Using a set of GWP as basis,
%\be
%\ket{\phi_n} = \sum_\alpha \ket{\alpha} C_{\alpha n}
%\ee
%\be
%H_{\beta \alpha} C_{\alpha n} = E_n S_{\beta \alpha} C_{\alpha n}
%\ee
%
%For the potential energy operator
%\be
%x_{\beta \alpha} = \braket{\beta | x|\alpha}
%\ee
%%if we diagonalize $x$ by
%%\be
%%x = U^\dag \Lambda U
%%\ee
%%\textbf{This is wrong!}
%We should do
%\be
%x \ket{x_n} = x_n \ket{x_n}
%\ee
%say
%\be
%\ket{x_n} = C_{\alpha n} \ket{\alpha}
%\ee
%then
%\be
%x C =  S C X
%\ee

%The local diabatic picture solves the singularity problem, how to address the exponential scaling?

%our ansatz reads
%\be
%\ket{\Psi} = \sum_\alpha \del{\sum_n C_{n\alpha} \ket{\psi_n(\bf R_\alpha)} } \ket{\bf R_\alpha}
%\label{eq:101}
%\ee
%where $\Psi$ is the molecular state, $\hat{R}_\mu \ket{R_\alpha} = R_\mu \ket{\bf R^\alpha}$

Inserting \eq{eq:100} into the \tdse\ for the molecule $i \pd{\Psi}{t} = \del{H_\text{BO} + \hat{T}_\text{n}} \Psi$ and left multiply $\bra{\psi_\beta(\bf R_m)}\bra{\bf R_m}$ yields

\be
i \dot{C}_{m \beta}(t)
%\phi_\alpha(\bf r; \bf R_n) \chi_n(\bf R; \bf R_n, \bf P_n)
= E_\beta(\bf R_m)C_{m \beta}(t)+  T_{mn}A_{m\beta, n\alpha} C_{n\alpha}
\label{eq:main}
\ee
where we have made use of 
\be 
H_\text{BO}(\bf R) \ket{\psi_n(\bf R_\alpha)} \ket{\bf R_\alpha} = E_n(\bf R_\alpha) \ket{\psi_n(\bf R_\alpha)} \ket{\bf R_\alpha} 
\label{eq:dvr}. 
\ee  
Here $T_{\beta \alpha} = \braket{\bf R_\beta | \hat{T}_\text{n} | \bf R_\alpha}_{\bf R}$ is the kinetic energy operator matrix elements and the electronic overlap matrix
\be
A_{m\beta, n\alpha} = \braket{\psi_{m}(\bf R_\beta) | \psi_n(\bf R_\alpha) }_{\bf r},
\ee
where $\braket{\cdots}_{\bf r}$ ($\braket{\cdots}_{\bf R}$) denotes the integration over electronic (nuclear) degrees of freedom.  In \eq{eq:dvr}, we have employed an approximation used in the discrete variable representation \cite{light2000} $\braket{\bf R_n| V(\bf R) | \bf R_n}  \approx V(\bf R_n) $.  

 \eq{eq:main} can also be derived by Dirac-Frenkel variational principle $\delta \int_{t_i}^{t_f} \dif t \braket{\Psi|i \pa_t - H | \Psi} = 0$.  
The first term in the right-hand side of \eq{eq:main} gives a dynamical phase depending on the APESs. 
The electronic overlap matrix $A$ appearing in the second term encodes all nonadiabatic effects and geometric phase. It plays a similar role to nonadiabatic coupling in the adiabatic representation. 
For any $n$, 
$
A_{n\beta, n\alpha} = \delta_{\beta \alpha}
$
due to orthonormality of the adiabatic electronic states. 
Also, it is  Hermitian
\be
A_{m\beta, n\alpha } = A_{ n\alpha, m\beta}^*
\ee
The overlap matrix is gauge-dependent, and transform as  $A'_{m\beta, n\alpha} =  A_{m\beta, n\alpha} e^{-i \del{\theta_n - \theta_m}}$ under the gauge transformation in \eq{eq:gauge}.   But consider a loop encircling a CI with geometries labeled by $1, 2, \cdots, N$,  the  Wilson loop  $W_\alpha =  A_{1 \alpha, 2 \alpha}  A_{2 \alpha, 3 \alpha}\cdots  A_{N \alpha, 1 \alpha}$ reflects the geometric phase and is  gauge invariant \cite{fomenko2009}. 

The expansion coefficients for different electronic states becomes decoupled in the limit where $A_{m\beta, n\alpha} \approx A_{mn}^{\alpha} \delta_{\beta \alpha}$. This implies that the ground electronic state at $\bf R$ and excited electronic state at $\bf R' \ne \bf R$ are orthogonal. In other words, let $\bf R' = \bf R + \Delta \bf R$, $\braket{\phi_g(\bf R) | \phi_e(\bf R + \Delta \bf R) } = \Delta \bf R \cdot \braket{\phi_g(\bf R) | \grad_{\bf R} | \phi_e(\bf R)} = 0$. Thus, this approximation is equivalent to neglecting the nonadiabatic couplings, the LDR is consistent with the adiabatic representation in this limit.  However, the geometric phase effects in the adiabatic dynamics is still captured by $A_{mn}^\alpha$.

\eq{eq:main} offers several advantages for nonadiabatic quantum dynamics simulations.
% of both the adiabatic and diabatic representations, yet without the singularities and
%In the adiabatic representation, the nonadibatic coupling diverges at the CI point. 
First, although we have employed the adiabatic electronic states, it does not contain any singularities because the nuclear kinetic energy operator does not operate on the electronic states. Therefore, in contrast to wave packet dynamics in the adiabatic representation, the singularity of the nonadiabatic coupling at the CI point will not affect the simulations. In fact, the CI configuration can be safely  included in our basis set. 
% which does not affect the simulations. 
%This is in contrast to the adiabatic representation where the CI cannot be included.
%
Moreover, \eq{eq:main} can be used for an arbitrary gauge. It does not require the electronic wavefunction to be continuous with respect to the nuclear geometry because there is no term involving the nuclear derivative of the electronic wavefunction. This is advantages for quantum dynamics simulations as electronic structure calculations at different geometries can be directly used without further smoothing procedure. 

In practice, the nuclear kinetic energy matrix elements can be analytically calculated within the primitive basis and then transformed into the position eigenstates. The adiabatic electronic states can be computed with well-established electronic structure methods. 
Any nuclear basis set can be used. Depending on the nuclear degree of freedom, different basis may be more convenient and require less basis functions to converge than the other. For example, a Fourier basis may be convenient for rotational dynamics (e.g. in isomerization) %angular coordinates 
whereas local Gaussian bases including  coherent states may be useful for describing vibrational dynamics.  When  a localized basis set is parameterized by a configuration (e.g. the Gaussian basis is parameterized by a center), it is possible to directly employ the adiabatic electronic states defined at the corresponding configurations,  as in  the moving crude adiabatic representation \cite{ryabinkin2017}. This can also avoid the singular nonadiabatic couplings, but the vibronic basis set are not orthogonal and may encounter singular overlap matrix. 
%\red{read the paper and see whether there are further disadvantages} 
%In the limit of a single Gaussian basis, the position eigenvalue is simply the center, our ansatz reduces to the 

%\subsection{Observables}
The electronic and nuclear operators in the LDR can be constructed from 
the resolution of identity of the joint electron-nuclear space, 
$
\mc{I} = \sum_{n, \alpha} \ket{\phi_\alpha(\bf R_n), \bf R_n}   \bra{\phi_\alpha(\bf R_n), \bf R_n}.  
$ 
This follows from the orthonormality of the basis set and can be verified by $\mc{I} \ket{\Psi} = \ket{\Psi}$ for any vibronic state $\Psi$.  For a joint electron-nuclear operator $O(\bf r, \bf R)$, 
\be 
\begin{split} 
O &= \mc{I} O \mc{I} \\
&= \sum_{m,n} \sum_{\beta, \alpha} O_{m\beta, n\alpha} \ket{\phi_{\beta}(\bf R_m), \bf R_m} \bra{\phi_\alpha(\bf R_n), \bf R_n} \\
\end{split}
\ee 
where $O_{m\beta, n\alpha} = \braket{\phi_\beta(\bf R_m), \bf R_m | O | \phi_\alpha(\bf R_n), \bf R_n}$. 
The electronic and nuclear reduced density matrices read, respectively,  
$
\rho_\text{e}(t) = \sum_n C_{n\alpha}(t) C_{n\beta}^*(t) \ket{\phi_\alpha(\bf R_n)} \bra{\phi_\beta(\bf R_n)} 
$ 
and 
$
\rho_\text{n}(t) = \sum_{n,m} \sum_{\beta, \alpha} C_{m\beta}^* A_{m\beta, n\alpha} C_{n\alpha} \ket{\bf R_n} \bra{\bf R_m} 
$. For a purely  electronic (nuclear) observable $O$, the expectation value can be simplified as 
$
\braket{O(t)}  = \sum_{n, \beta \alpha}C_{n\beta}^* O^{n}_{\beta \alpha} C_{n\alpha}
$
($
\braket{O(t)}  = \sum_{m,n} \sum_{ \beta, \alpha} C^*_{m\beta} O_{mn} A_{m\beta, n\alpha} C_{n\alpha}  
$).

\section{Application to conical intersection model}
\begin{figure*}[hbtp]
	\includegraphics[width=\textwidth]{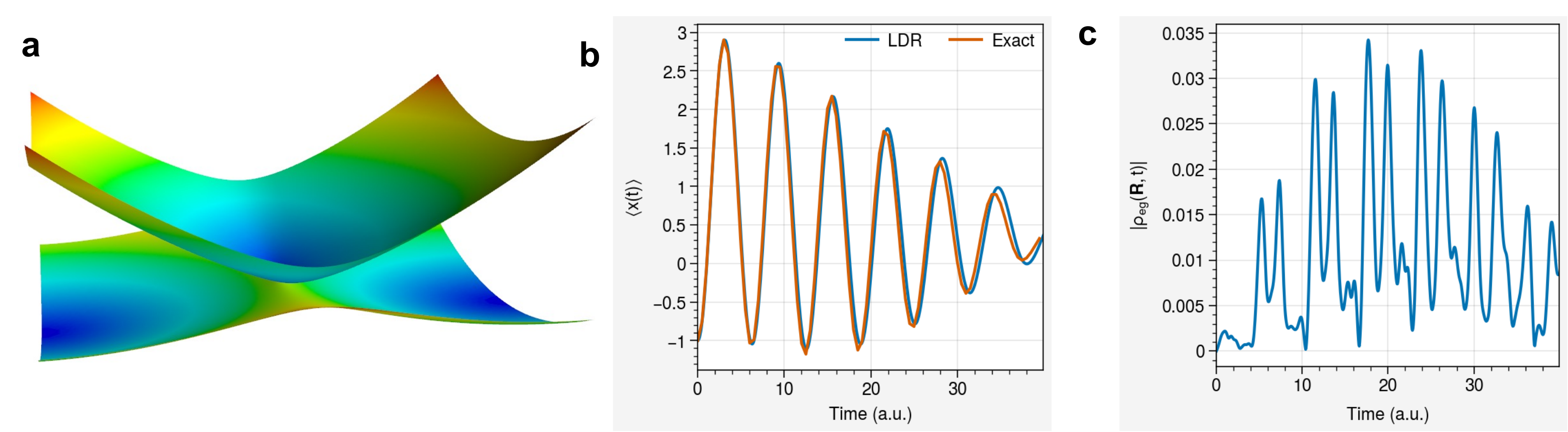}
	\caption{Wave packet dynamics of a two-dimensional conical intersection model. (a) Adiabatic potential energy surfacs. (b) Expectation value of the position operator $x$ reflecting the wavepacket motion. (c) Electronic coherence at a point near conical intersection. The parameters of the model are chosen as (atomic units) $\kappa = 1, \lambda = 0.2, \Delta = 1$.}
	\label{fig:dynamics}
\end{figure*}

We demonstrate the utility of the LDR for nonadiabatic wavepacket dyn amics by a two-state two-dimensional conical intersection model. The Hamiltonian reads 
\be
H = \sum_{\sigma = x, y} \del{a_{\sigma}^\dag a_{\sigma}  + \half }  - \kappa x \sigma_z + \lambda y \sigma_x + \Delta \sigma^+\sigma^- %\del{\ket{0}\bra{1} + \ket{1}\bra{0}} 
\ee 
where $\sigma_x = \ket{0}\bra{1} + \ket{1}\bra{0}$ and $\sigma_z = \ket{1}\bra{1} - \ket{0}\bra{0}$, $\ket{0}, \ket{1}$ are two diabatic states. The adiabatic potential energy surfaces are depicted in \cref{fig:dynamics}a, with a CI located at $(x, y) = (0, 0)$. The tuning $x$ tunes the energy gap and the coupling $y$ induces electronic transitions. Quantum dynamics of the diabatic vibronic Hamiltonian can be modeled exactly by a split-operator method with the nuclear wave packets associated with each electronic state represented in a numerical grid \cite{kosloff1988}. 
Initially, the molecular state is $\Psi_0 = {\pi}^{-1/2}  e^{- \half (x+1)^2 - \half y^2} \proj{1} $, arising from a vertical excitation from the ground state.  
%{we want to calculate the electronic coherence and geometric phase.}

%the nuclear density reads 
%$
% \rho(\bf R, t) = \braket{\bf R|\rho_\text{n}(t)|\bf R} 
% %= \sum_{n,m} \sum_{\beta, \alpha} C_{m\beta}^* A_{m\beta, n\alpha} C_{n\alpha} \chi_n(\bf R) \chi^*_m(\bf R)%\ket{\bf R_n} \bra{\bf R_m} 
%$ 

%If we want to know the nuclear wavepacket associated with the electronic ground state, the operator 
%\be
%P = \sum_n \proj{\phi_0(\bf R_n)} 
%\ee 

\begin{figure}[hbtp]
	\includegraphics[width=0.5\textwidth]{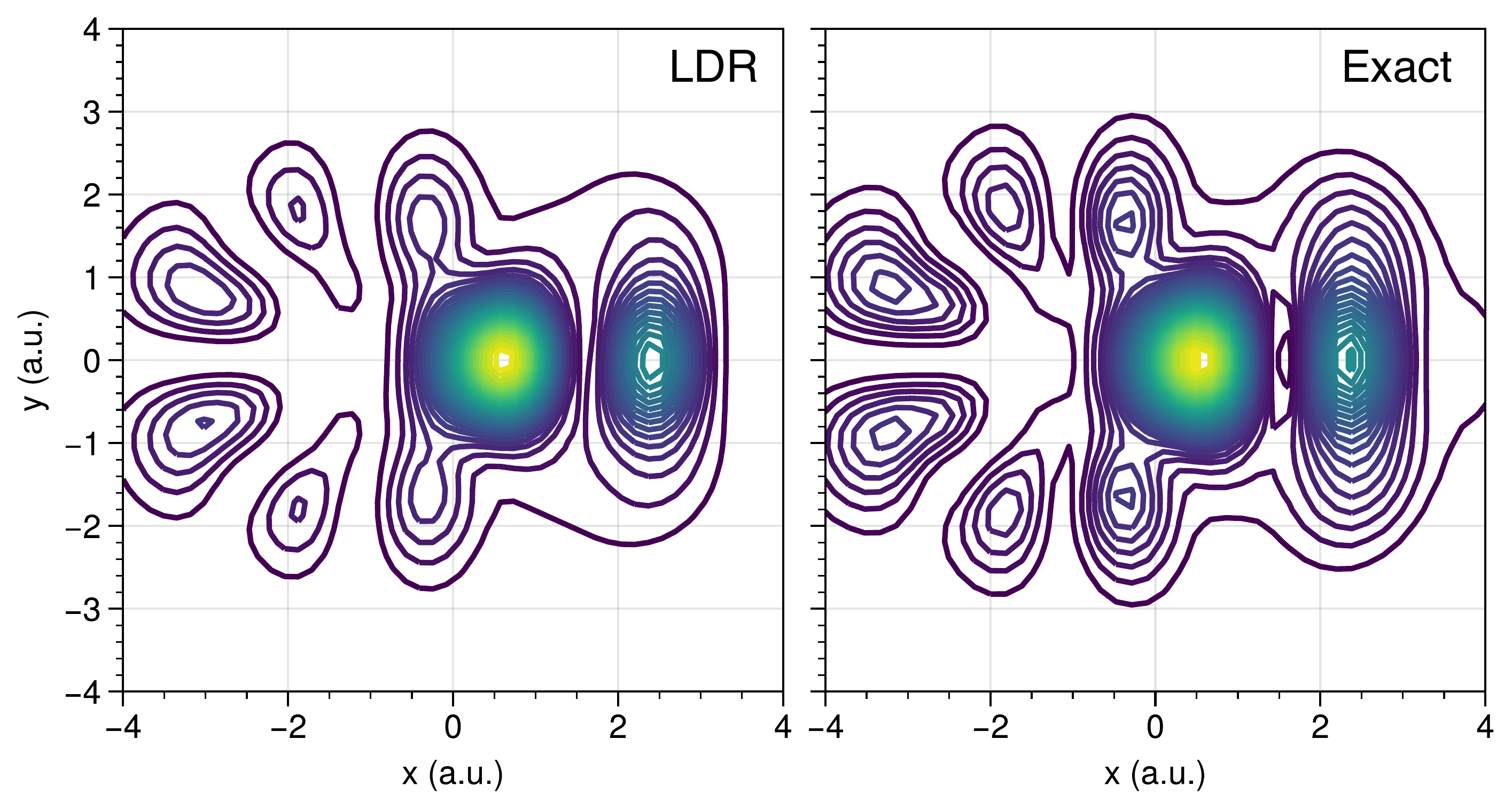}
	\caption{Geometric phase in the total nuclear probability density at $t = 40$ a.u.. The geometric phase is reflected in the nodal line along $y = 0.$}
	\label{fig:wavepacket}
\end{figure}

%\section{Computational details}

Here we use coherent states as the primitive nuclear basis set, whereby the matrix elements of position and momentum operators can be analytically calculated.  \eq{eq:main} is solved with the fourth-order Runge-Kutta integrator. \cref{fig:dynamics}b shows the expectation value of the tuning mode, which reflects the wave packet motion. The oscillations reflect multiple passages through the CI. The electronic coherence are reflected in the off-diagonal elements of $C(\bf R) = \ket{g(\bf R)} \bra{e(\bf R)}$. \cref{fig:dynamics}c shows the electronic coherence at a point close to the CI. The geometric phase effects lead to a nodal line in the nuclear probability density, see \fig{fig:wavepacket}.  Here the LDR employs adiabatic states, obtained by diaganolizing the electronic Hamiltonian at eigenvalues of the position operators, with an random phase assigned. The CI point is included in our basis set. The results simulated with LDR are in excellent agreement with the exact quantum results.

\section{summary}
To summarize, we have developed a locally diabatic representation for the correlated electron-nuclear quantum dynamics around conical intersections. It employs the adiabatic electronic states without the diabatization and without introducing the nonadiabatic coupling into the dynamical equations, thus avoiding the singularity associated with conical intersections. By a two-dimensional CI model, we have demonstrated that it can describe accurately the nuclear wave packet passing through a CI, with results in excellent agreement with exact quantum dynamics. The LDR can be useful for problems involving degenerate electronic states. This occurs not only due to conical intersections but also due to spin, for example, in intersystem crossing. 
This paves the way for ab initio nonadiabatic wave packet dynamics and molecular spectroscopy simulations. 
%We have demonstrated its utility in simulating conical intersection dynamics and showed that it shows an excellent agreement with quantum dynamics with 

The locally diabatic representation can be  generalized to describe any fiber bundle structure \cite{mead1982, cohen2019}, which  are recognized to be important in many areas of quantum physics, molecular science, and optics. The base space of a bundle will be described within a finite basis representation.  After transforming to the localized basis $\set{X_n}$, an adiabatic representation ${F\del{X_n}}$ is used for the fiber associated with the basis $X_n$.  That is, the full space is covered with a set of  product spaces  ${F\del{X_n}} \times X_n$. The transformation to localized basis set ensures that the spaces are orthogonal to each other. The local diabatic representation is in essence a local trivialization process whereas the diabatization is a global trivialization, which is only possible if the bundle is topologically trivial.

The computational cost of solving \eq{eq:main}  scales quadratically with the size of basis set rather than with the number of nuclear degrees of freedom. Although with a direct product basis set, the number of basis functions scales exponentially with  system size. Nevertheless, for bound potential energy surface, a small number of randomly distributed high-dimensional Gaussian bases can be used \cite{garashchuk2001}, thus alleviating the increase of basis functions at high-dimensional quantum dynamics. The computational efficiency may be further improved by adopting moving basis set such as trajectory-guided basis \cite{gu2016b}. These directions will be explored in the future.

\bibliography{../../cavity,../../optics,../../qchem,../../topology}

\end{document}